\documentclass[aps, prd, groupedaddress, floatfix]{revtex4-2}
\usepackage{color}
\usepackage{amssymb}
\usepackage[normalem]{ulem}
\usepackage{graphicx,color}
\usepackage{amsmath}
\usepackage{subfigure}
\usepackage[normalem]{ulem}
\usepackage{booktabs}
\usepackage{xcolor}

\colorlet{darkgreen}{green!50!black}
\colorlet{brightyellow}{yellow!75!red}
\colorlet{orange}{red!50!yellow}
\colorlet{darkblue}{blue!60!black}
\colorlet{darkred}{red!80!black}

\usepackage{soul}

\usepackage{mathtools}
\usepackage{mathrsfs}
\usepackage{bm}
\usepackage{relsize}
\usepackage{indentfirst}

\usepackage{xcolor}

\usepackage{amssymb, amsmath, multirow, epsfig, graphics, color}

\usepackage{epsfig}
\usepackage{graphicx,amsmath}
\def\be{\begin{eqnarray} &&}

\def\ee{\end{eqnarray}}

\newcommand\ba{\begin{eqnarray}}
\newcommand\ea{\end{eqnarray}}

\newcommand{\bas}{\begin{eqnarray*}}
\newcommand{\eas}{\end{eqnarray*}}

\newcommand{\bno}{\begin{eqnarray*}}
\newcommand{\eno}{\end{eqnarray*}}

\def\sl
\usepackage{hyperref}
\usepackage{url}
\usepackage{hyperref}
\hypersetup{
colorlinks=true,  
linkcolor=blue,   
filecolor=magenta,
urlcolor=blue,    
citecolor=blue,   
} 
\begin{document}
\title{A Dark Matter Model with Quadratic Equation of State:\\ Background Evolution and Structure Formation}

\author{
K. Rezazadeh$^{1}$\footnote{kazem.rezazadeh@ipm.ir},
E. Yusofi$^{1,2}$\footnote{eyusofi@ipm.ir},
and A. Talebian$^{1}$\footnote{talebian@ipm.ir},
}

\affiliation{
$^{1}$\small{School of Astronomy, Institute for Research in Fundamental Sciences(IPM), P. O. Box 19395-5531, Tehran, Iran}\\
$^{2}$\small{Innovation and Management Research Center, AA. C., Islamic Azad University, Amol, Iran}
}

\date{\today}

\begin{abstract}
We propose that dark matter (DM) possesses a quadratic equation of state, which becomes significant at high densities, altering the Universe's evolution during its early stages. We derive the modified background evolution equations for the Hubble parameter $H(z)$ and the DM density parameter $\Omega_{\text{dm}}(z)$. We then perturb the governing equations to study the linear growth of matter fluctuations, computing the observable growth factor $f\sigma_8(z)$. Finally, we compare the model with the latest cosmological data, including Hubble parameter $H(z)$ measurements, and growth factor $f\sigma_8(z)$ data, up to $z=3$. Our results indicate that the quadratic model, while remaining consistent with background observations, offers a distinct imprint on the growth of structure, providing not only a new phenomenological avenue to address cosmological tensions but also shedding light on the nature of DM.
\\
\noindent \hspace{0.35cm} 
\noindent \hspace{0.35cm} 
\end{abstract}

\maketitle

\section{Introduction}
\label{sec:introduction}

The $\Lambda$CDM model stands as the cornerstone of modern cosmology, providing a remarkably successful description of the Universe's evolution from its primordial perturbations to its large-scale structure (LSS) today \cite{Planck:2018vyg, SupernovaSearchTeam:1998fmf, SupernovaCosmologyProject:1998vns}. Although striking, the success is accompanied by persistent tensions and unresolved theoretical questions. Notably, the Hubble tension \cite{Riess:2020fzl, DiValentino:2021izs}, a $\sim 5\sigma$ discrepancy between the early-Universe measurement of $H_0$ from the Cosmic Microwave Background (CMB) \cite{Planck:2018vyg} and late-Universe distance ladder measurements \cite{Riess:2020fzl}, challenges the standard cosmological paradigm, likely pointing to exciting new physics yet to be discovered \cite{Riess2024, CosmoVerse2025}.

Similarly, the $S_8$ tension, concerning the amplitude of matter fluctuations, persists between CMB inferences and weak gravitational lensing surveys \cite{Joachimi:2020abi, KiDS:2020suj}. Furthermore, recent observations from the James Webb Space Telescope (JWST) have uncovered an unexpectedly high abundance of massive galaxies at very high redshifts ($z \gtrsim 10$) \cite{Labbe:2022ahb}, suggesting potentially more efficient structure formation in the early Universe than predicted by standard $\Lambda$CDM \cite{Boylan-Kolchin:2022kae}. Recent analyses further suggest the $S_8$ tension may be confined to late-universe physics ($z \leq 2$) \cite{Lin:2023uux}. Furthermore, the analysis of the DESI collaboration reports compelling evidence for evolving dark energy, including potential phantom divide crossing ($w = -1$), conflicting with a pure cosmological constant scenario \cite{Adame2025, DESI:2024mwx}.

These tensions motivate the investigation of new physics in DM. Several compelling candidates could explain the nature of DM, including isocurvature perturbations during inflation \cite{Firouzjahi:2021lov}, axions \cite{Briaud:2023pky}, primordial black holes \cite{Hooshangi:2022lao, Talebian:2022cwk, HosseiniMansoori:2023mqh}, and even axion-like particles produced in the radiation-dominated era \cite{Talebian:2023lkk}. One conceivable avenue is to consider that DM may have non-trivial properties, moving beyond its traditional description as a cold, collisionless, and pressureless ($P_{\rm dm}=0$) fluid. For instance, models involving gravitational particle production \cite{Lima:1995xz, Lima:2015mca, Safari2022} treat the Universe as an open thermodynamic system, introducing an effective pressure that modifies the continuity equation. This approach can provide a better fit to both background and perturbation data compared to $\Lambda$CDM. In a different approach, \cite{Davari:2023tam} investigated DM with a constant and a dynamically evolving equation of state within the spherical collapse framework. They demonstrated that such models significantly impact the linear overdensity threshold for collapse and the abundance of high-redshift halos. It potentially enhances early structure formation to levels more consistent with JWST observations. For other scenarios as alternatives to the standard cold DM, see e.g. \cite{Burrage:2016yjm, OHare:2018ayv, Burrage:2018zuj, Kumar:2019gfl, Ashoorioon:2023jwf, Kading:2023hdb, Montani2024, Kumar:2025etf, Yang2025, Chen2025, Jusufi:2025hte, Braglia:2025gdo}.

Concurrently, there has been some interest in cosmological fluids with non-linear equations of state \cite{Ananda:2005xp, Chavanis:2012pd, Chavanis:2012kla, Chavanis:2014lra, Mohammadi:2023idz, Shahriar:2024vqx, Moshafi2024}. A quadratic EoS offers a phenomenologically rich framework. The quadratic term becomes significant at high densities (early times) and negligible at lower densities (late times), offering a unique redshift-dependent effect \cite{Moshafi2024}. This has been explored primarily in the context of dark energy to achieve phenomena like phantom divide crossing \cite{Nojiri2005, Shahriar:2024vqx} and to address the Hubble tension \cite{Moshafi2024, Mohammadi:2023idz}. Dynamical systems analysis reveals that such models can feature stable attractors and act as cosmic energy sources or sinks, governed by the quadratic parameter \cite{Mohammadi:2023idz}.

In this work, we propose the Quadratic Density-dependent Dark Matter (QDDM), which is characterized by the following quadratic equation of state
\begin{equation}
\label{eq:main_eos}
P_{\rm dm} = \alpha \, \rho_{\rm dm} + \beta \, \frac{\rho_{\rm dm}^{2}}{\rho_0} \, ,
\end{equation}
where $P_{\rm dm}$ and $\rho_{\rm dm}$ are the pressure and energy density of the DM component, respectively. Here, $\rho_0$ indicates the current critical density and serves to normalize the quadratic term. The parameters $\alpha$ and $\beta$ are dimensionless constants. This ansatz is phenomenologically rich and can be motivated by certain field theories where the pressure is a non-linear function of the density \cite{Ananda:2005xp}. In the conventional DM models, both parameters $\alpha$ and $\beta$ are zero.

The quadratic term, proportional to $\rho_{\rm dm}^2$, is designed to become significant only in high-density regimes, such as the early universe, while being negligible at late times, offering a unique redshift-dependent effect that differs from previously studied models \cite{Pace2020, Davari:2023tam}. This introduces a new mechanism to alter the growth history of structures, particularly at intermediate and low redshifts ($0 < z < 3$), without drastically violating well-constrained late-time cosmology.

We aim to perform an analysis of this model. We will derive the modified background evolution equations for the Hubble parameter $H(z)$ and the DM density parameter $\Omega_{\rm dm}(z)$. We will then derive the equations governing the linear growth of matter perturbations and compute the observable growth factor $f\sigma_{8}(z)$. Finally, we will compare the model with the latest cosmological data, including Hubble parameter $H(z)$ measurements, and growth factor $f\sigma_{8}(z)$ data, and perform a comparison with the standard $\Lambda$CDM model.

The paper is structured as follows: In Section \ref{sec:model}, we present the theoretical framework of our QDDM model, deriving the background evolution equations. Section \ref{sec:perturbations} is devoted to the study of the linear perturbations in the setup of QDDM model. Then, in Sec. \ref{sec:growth_factor}, we compute the growth factor in our model, and compare our findings with the $\Lambda$CDM model. Finally, we summarize our conclusions and discuss future outlooks in Section \ref{sec:conclusions}.

\section{Background Dynamics}
\label{sec:model}

We begin by considering a homogeneous and isotropic Universe described by the flat Friedmann-Lemaître-Robertson-walker (FLRW) metric. The evolution of the cosmic scale factor $a(t)$ is governed by the Friedmann equations
\begin{align}
H^2 &\equiv \left(\frac{\dot{a}}{a}\right)^2 = \frac{8\pi G}{3} \rho_{\text{tot}} \, , \label{eq:Fried1} \\
\frac{\ddot{a}}{a} &= -\frac{4\pi G}{3} (\rho_{\text{tot}} + 3P_{\text{tot}}) \, , \label{eq:Fried2}
\end{align}
where the dots denote the derivatives with respect to the cosmic time and $H$, $\rho_{\text{tot}}$, and $P_{\text{tot}}$ are the Hubble rate, the total energy density, and the total pressure, respectively. The current value of the Hubble rate $H_0$ and the critical density $\rho_0$ are related via \eqref{eq:Fried1}.

In the standard $\Lambda$CDM paradigm, the DM component is assumed to be pressureless ($P_{\text{dm}}=0$). In this work, we propose (\ref{eq:main_eos}) as EoS for DM. This EoS reduces to the standard pressureless case for $\alpha=0$ and $\beta=0$. The evolution of the DM energy density is determined by the energy conservation equation
\begin{equation}
\label{eq:cons_full}
\dot{\rho}_{\text{dm}} + 3H\rho_{\text{dm}}(1 + \alpha + \beta \frac{\rho_{\text{dm}}}{\rho_0}) = 0 \, .
\end{equation}
This equation can be integrated to find the evolution of the DM density $\rho_{\text{dm}}(z)$ in which $z$ represents the cosmological redshift.
The solution of \eqref{eq:cons_full} is given by
\begin{equation}
\label{eq:rho_sol_general}
\rho_{\text{dm}}(z) = \rho_{\text{dm0}} F(z) \, ;
\hspace{1.5cm}
F(z)\equiv
\frac{(1 + \alpha)}{(1 + \alpha + \beta \, \Omega_{\text{dm0}})\,(1+z)^{-3(1+\alpha)} - \beta \, \Omega_{\text{dm0}}} \, .
\end{equation}
Here $\rho_{\text{dm0}}$ is the present-day ($z=0$) DM density and $\Omega_{\text{dm0}}$ is the current DM fractional energy density, $\Omega_{\text{dm0}} \equiv \rho_{\text{dm0}}/\rho_0$. The dimensionless density function $F(z)$ describes the deviation from the standard $\Lambda$CDM evolution. For the case $\alpha =\beta =0$, this reduce to the $\Lambda$CDM case, $F(z)=(1+z)^3$.

\begin{figure}[ht]
\centering
\includegraphics[width=7.6 cm]{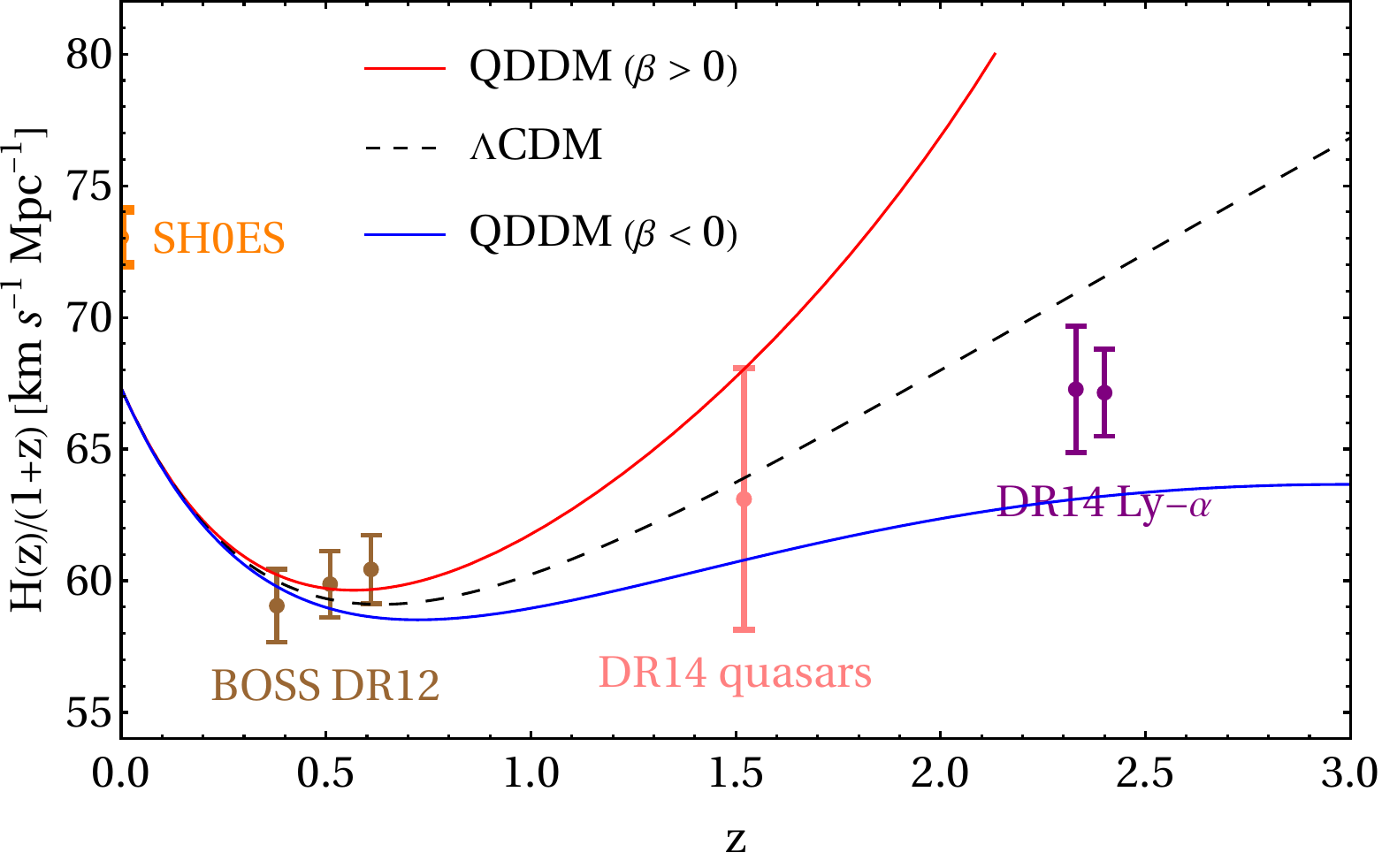}
\hspace{1.5cm}
\includegraphics[width=7.6 cm]{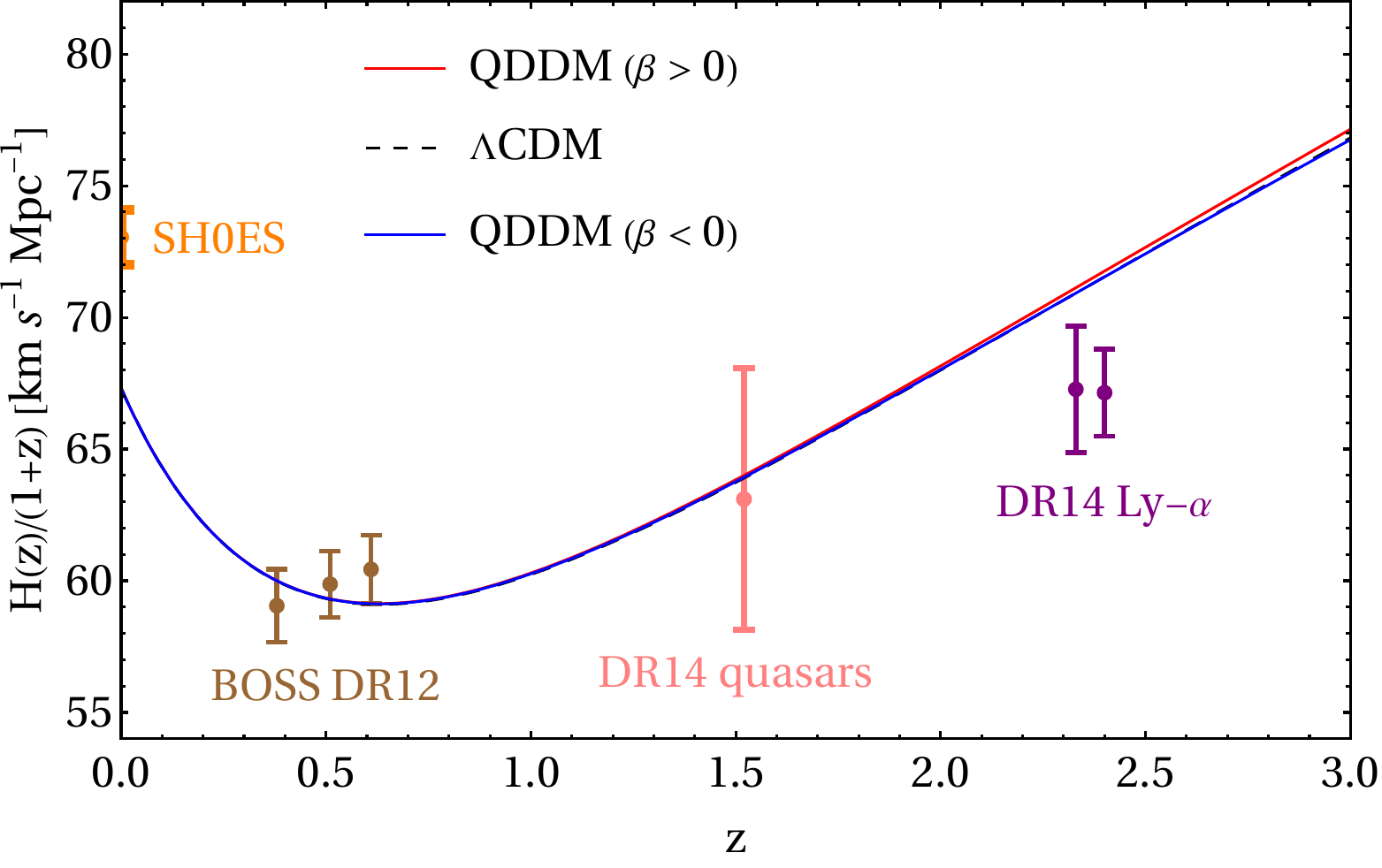}
\caption{\footnotesize Evolution of scaled Hubble parameter in terms of the cosmological redshift for our model with $\beta > 0$ (red curve) and with $\beta < 0$ (blue curve), compared to the $\Lambda$CDM model (black dashed curve). The left plot is drawn with $\{\alpha,\left|\beta\right|\}=\{10^{-3},10^{-2}\}$, the right one with $\{\alpha,\left|\beta\right|\}=\{10^{-3},10^{-4}\}$. To generate these plots, we have used the cosmological parameters $\{h,\Omega_{b0},\Omega_{\text{dm0}},\Omega_{r0}\}=\{0.6732,0.04939,0.265,9.267\times10^{-5}\}$.}
\label{fig:H}
\end{figure}

\begin{figure}[ht]
\centering
\includegraphics[width=8 cm]{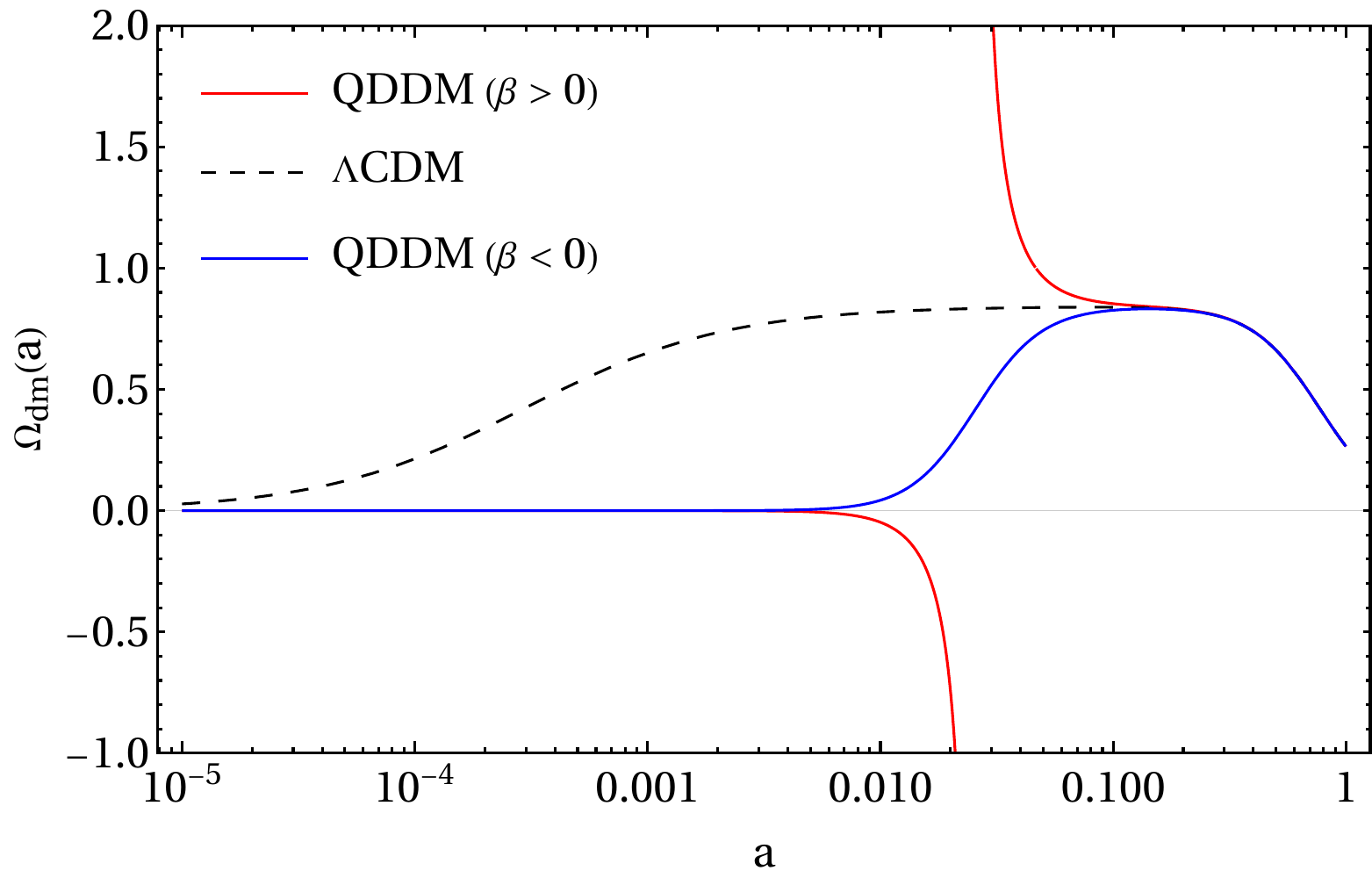}
\caption{\footnotesize Evolution of DM density parameter in terms of scale factor for our model with $\beta = 10^{-4}$ (red curve) and with $\beta < 0$ (blue curve), compared to the $\Lambda$CDM model (black dashed curve). To draw this plot, we used $\{\alpha,\left|\beta\right|\}=\{10^{-3},10^{-4}\}$. The values of the other cosmological parameters are the same as those used in Fig. \ref{fig:H}.}
\label{fig:Omega}
\end{figure}

The total energy density of the Universe is the sum of its components: dark energy ($\Lambda$), DM, baryons ($b$), and radiation ($r$),
\begin{equation}
\rho_{\text{tot}} = \rho_{\text{dm}} + \rho_{b} + \rho_{r} + \rho_{\Lambda} \, .
\end{equation}
Substituting thin into the first Friedmann equation (Eq.~\eqref{eq:Fried1}) and expressing the fractional densities in terms of their present-day density parameters $\Omega_{i0} = \rho_{i0} / \rho_0$, we obtain the Hubble parameter for our model
\begin{equation}
\label{eq:H_z}
H(z) = H_0 \sqrt{ \Omega_{\text{dm0}} F(z) + \Omega_{b0}(1+z)^3 + \Omega_{r0}(1+z)^4 + \Omega_{\Lambda 0} } \, .
\end{equation}
The evolution of the dark matter density parameter is given by
\begin{equation}
\label{Omega_dm_z}
\Omega_{\text{dm}}(z) = \frac{\Omega_{\text{dm0}} F(z)}{\left( H(z) / H_0 \right)^2} \, .
\end{equation}
As seen from \eqref{eq:H_z} and \eqref{Omega_dm_z}, the background dynamics of the Universe are affected by the non-trivial function $F(z)$. In Figs. \ref{fig:H} and \ref{fig:Omega}, we have illustrated the evolution of $H(z)$ and $\Omega_{\text{dm}}(z)$ and compare our model with the standard $\Lambda$CDM case.

Fig. \ref{fig:H} depicts the evolution of the scaled Hubble rate in terms of redshift for different cosmological scenarios. The black dashed curve represents the standard $\Lambda$CDM model. Our model introduces a deviation governed by the sign and magnitude of the parameter $\beta$. A positive value of $\beta$ (red curve) signifies a positive quadratic pressure term, which acts as an effective additional pressure, resisting gravitational collapse. This leads to a more rapid expansion at early times (high density) compared to $\Lambda$CDM. Conversely, a negative $\beta$ (blue curve), implies a negative pressure contribution, enhancing the gravitational attraction within the dark matter fluid and resulting in a slower early-time expansion. At late times, as the density $\rho_{\rm dm}$ decreases, the quadratic term becomes subdominant, and all models converge to the accelerated expansion driven by the cosmological constant.

This modified expansion history directly impacts the evolution of the DM density parameter $\Omega_{\rm dm}(z)$, as shown in Fig. \ref{fig:Omega}. In the $\Lambda$CDM scenario, we have $\Omega_{\rm dm} \propto (1+z)^3$, which is shown by the black dashed curves. Our model deviates from this scaling. For $\beta > 0$ (red curve), the effective pressure suppresses the energy density relative to $\Lambda$CDM at high redshifts, causing $\Omega_{\rm dm}(z)$ to be lower. For $\beta < 0$ (blue curve), the negative pressure enhances the clustering, leading to a higher $\Omega_{\rm dm}(z)$ at early times. This divergence in the matter density history is a key distinguishing feature that can be probed by combinations of CMB, BAO, and SNIa data.

\section{Linear Perturbations}
\label{sec:perturbations}

In this section, we derive the equations governing the linear evolution of matter density perturbations for the QDDM model. We work in the Newtonian gauge, which is well-suited for studying sub-horizon perturbations in the Universe. For a barotropic fluid, indicated by the subscript $i$, we consider an equation of state $P_i = P_i(\rho_i)$. In addition, we assume $\bar{\rho}_i$ to be the background energy density and $\delta \rho_i$ to be its perturbations in the Newtonian gauge. The evolution of the density contrast, $\delta_i = \delta\rho_i / \bar{\rho}_i$, and the divergence of the peculiar velocity, $\theta_i = \vec{\nabla} \cdot \vec{v}_i$, are given by the continuity and the Euler equations in Fourier space.

For the $i$-th fluid in the Newtonian gauge, the equations governing the mode functions (in Fourier space) with wave-number $k$ are given by \cite{Lima:1996at, Hwang:1997xt, Hwang:2005xt, Abramo:2008ip, Fahimi:2018pcr}
\begin{align}
& \dot{\delta}_{i}+3H\left(c_{s,i}^{2}-w_{i}\right)\text{\ensuremath{\delta}}_{i}+\frac{\left(w_{i}+1\right)}{a}\theta_{i}=0 \, ,
\label{delta_i_dot}
\\
& \dot{\theta}_{i}+H\text{\ensuremath{\theta}}_{i}-\frac{k^{2}c_{s,i}^{2}}{a\left(w_{i}+1\right)}\text{\ensuremath{\delta}}_{i}+\frac{3}{2}aH^{2}\underset{j}{\sum}\left(3c_{s,j}^{2}+1\right)\text{\ensuremath{\Omega}}_{j}\text{\ensuremath{\delta}}_{j}=0 \, .
\label{theta_i_dot}
\end{align}
In these equations, $c_{s,i}^{2}\equiv\delta P_{i}/\delta\rho_{i}$ refers to the effective velocity of the perturbations for each fluid and $w_i \equiv P_i/\rho_i$. We assume that the evolution of the perturbations is adiabatic, so that we can take the effective velocity of the perturbations to be equal to its adiabatic velocity up to a good approximation, so $c_{s,i}^{2}\approx\dot{{P}}_{i}/\dot{{\rho_i}}$. In \cite{Rezazadeh:2020zrd}, the authors have shown explicitly that this approximation is thoroughly valid in their model in the linear regime of perturbations. We expect this approximation to be valid in our scenario too.

It is convenient to rewrite Eqs. \eqref{delta_i_dot} and \eqref{theta_i_dot} in terms of the scale factor as follows
\begin{align}
& \delta_{i}'+\frac{\left(w_{i}+1\right)}{a}\tilde{\theta}_{i}+\frac{3}{a}\left(c_{s,i}^{2}-w_{i}\right)\text{\ensuremath{\delta}}_{i}+=0 \, ,
\label{delta_i_prime}
\\
& \tilde{\theta}_{i}'+\left(\frac{2}{a}+\frac{H'}{H}\right)\tilde{\text{\ensuremath{\theta}}}_{i}-\frac{k^{2}c_{s,i}^{2}}{a^{3}H^{2}\left(w_{i}+1\right)}\text{\ensuremath{\delta}}_{i}+\frac{3}{2a}\underset{j}{\sum}\left(3c_{s,j}^{2}+1\right)\text{\ensuremath{\Omega}}_{j}\text{\ensuremath{\delta}}_{j}=0 \, ,
\label{thetat_i_prime}
\end{align}
where the prime denotes the derivative with respect to the scale factor, and we have also defined the dimensionless quantity $\tilde{\theta}\equiv\theta/(aH)$.

In the QDDM model, the parameters of the equation of state and the speed of sound of DM are not constant, but are given by the following relations
\begin{align}
w_{\text{dm}} &= \alpha + \beta \, \Omega_{\rm dm0} F(z) \, ,
\label{eq:weff_pert}
\\
{c_s^2}_{,\text{dm}} &= \alpha + 2\beta \, \Omega_{\rm dm0} F(z) \, . \label{eq:cs2_pert}
\end{align}
These expressions are redshift-dependent due to the quadratic term in the pressure relation \eqref{eq:main_eos}. At high redshifts, where $\rho_{\text{dm}}$ is large, both $w_{\text{dm}}$ and $c_{s,\text{dm}}^2$ deviate significantly from their standard values, impacting the growth of perturbations. For $\alpha > 0$, thermodynamic stability ($c_{s,\mathrm{dm}}^{2}\geq0$) requires $\beta < 0$ \cite{Shahriar:2024vqx}.

Since we set the initial conditions from an epoch deep inside the matter-dominated era, we will neglect the contribution of the radiation fluctuations in the evolution of matter perturbations. Utilizing Eqs. \eqref{delta_i_prime} and \eqref{thetat_i_prime} for the baryon and dark matter components in our model, we get
\begin{align}
& \text{\ensuremath{\delta}}_{b}'+\frac{\tilde{\theta}_{b}}{a}=0 \, ,
\label{delta_b_prime}
\\
& \text{\ensuremath{\tilde{\theta}}}_{b}'+\left(\frac{H'}{H}+\frac{2}{a}\right)\text{\ensuremath{\tilde{\theta}}}_{b}+\frac{3}{2a}\left[\text{\ensuremath{\Omega}}_{b}\text{\ensuremath{\delta}}_{b}+\left(3\text{\ensuremath{c_{s,\mathrm{dm}}^{2}}}+1\right)\text{\ensuremath{\Omega}}_{\mathrm{dm}}\text{\ensuremath{\delta}}_{\mathrm{dm}}\right]=0 \, ,
\label{theta_b_prime}
\\
& \text{\ensuremath{\delta}}_{\mathrm{dm}}'+\frac{3}{a}\left(c_{s,\mathrm{dm}}^{2}-w_{\mathrm{dm}}\right)\text{\ensuremath{\delta}}_{\mathrm{dm}}+\frac{\left(w_{\mathrm{dm}}+1\right)}{a}\text{\ensuremath{\tilde{\theta}_{\mathrm{dm}}}}=0 \, ,
\label{delta_dm_prime}
\\
& \text{\ensuremath{\tilde{\theta}}}_{\mathrm{dm}}'+\left(\frac{2}{a}+\frac{H'}{H}\right)\text{\ensuremath{\tilde{\theta}}}_{\mathrm{dm}}-\frac{k^{2}c_{s,\mathrm{dm}}^{2}}{a^{3}H^{2}\left(w_{\mathrm{dm}}+1\right)}\text{\ensuremath{\delta}}_{\mathrm{dm}}+\frac{3}{2a}\left[\text{\ensuremath{\Omega}}_{b}\text{\ensuremath{\delta}}_{b}+\left(3\text{\ensuremath{c_{s,\mathrm{dm}}^{2}}}+1\right)\text{\ensuremath{\Omega}}_{\mathrm{dm}}\text{\ensuremath{\delta}}_{\mathrm{dm}}\right]=0 \, .
\label{theta_dm_prime}
\end{align}
These equations are four coupled ordinary differential equations, and by solving them, we can determine the evolution of $\delta_b$ and $\text{\ensuremath{\delta}}_{\mathrm{dm}}$ in terms of the scale factor in our setup. Having the evolution of these quantities at hand enables us to compute the evolution of the matter density contrast, which is given by
\begin{equation}
\label{delta_m}
\delta_{m}=\frac{\rho_{b}\delta_{b}+\rho_{c}\delta_{c}}{\rho_{b}+\rho_{c}} \, .
\end{equation}

To solve the set of coupled linear differential equations given in Eqs. \eqref{delta_b_prime}-\eqref{theta_dm_prime}, it is necessary to specify the initial conditions. For this purpose, we choose a time that is deep inside the matter-dominated era and denote the scale factor of that epoch by $a_i$. At this time, the relation $\delta_{m}\approx C_{m}a$ holds with a good approximation. Here, $C_m$ is a proportional constant that can be taken as unity without any loss
of generality. As a result, we determine the initial conditions as follows
\begin{align}
& \text{\ensuremath{\delta}}_{b}(a_{i})=C_{b}a_{i} \, ,
\label{delta_b_i}
\\
& \tilde{\theta}_{b}(a_{i})=-C_{b}a_{i} \, ,
\label{theta_b_i}
\\
& \text{\ensuremath{\delta}}_{\mathrm{dm}}(a_{i})=\frac{C_{b}a_{i}\left(1-\Omega_{b}(a_{i})\right)}{3c_{s,\mathrm{dm}}^{2}(a_{i})\text{\ensuremath{\Omega}}_{\mathrm{dm}}(a_{i})+\text{\ensuremath{\Omega_{\mathrm{dm}}}}(a_{i})} \, ,
\label{delta_dm_i}
\\
& \tilde{\theta}_{dm}(a_{i})=-\frac{C_{b}a_{i}}{\text{\ensuremath{\Omega}}_{\mathrm{dm}}(a_{i})\left(3c_{s,\mathrm{dm}}^{2}(a_{i})+1\right)\left(w_{\mathrm{dm}}(ai)+1\right)}\Bigg\{3\text{\ensuremath{\Omega}}_{\mathrm{dm}}(a_{i})\left(w_{\mathrm{dm}}(a_{i})+1\right)
\nonumber
\\
& +c_{s,\mathrm{dm}}^{2}(a_{i})\left[2\tilde{k}^{2}a_{i}\left(\text{\ensuremath{\Omega}}_{b}(a_{i})-1\right)+9\left(w_{\mathrm{dm}}(a_{i})+1\right)\text{\ensuremath{\Omega}}_{\mathrm{dm}}(a_{i})\right]\Bigg\} \, .
\label{theta_dm_i}
\end{align}
The parameter $C_b$ in the above equations is defined as
\begin{equation}
\label{Cb}
C_{b}=\frac{\text{\ensuremath{\Omega}}_{\mathrm{dm}}(a_{i})\left(3c_{s,\mathrm{dm}}^{2}(a_{i})+1\right)\left(\text{\ensuremath{\Omega}}_{b0}+\Omega_{\mathrm{dm0}}\right)}{\Omega_{\mathrm{dm0}}\left(1-\text{\ensuremath{\Omega}}_{b}(a_{i})\right)+\text{\ensuremath{\Omega}}_{\mathrm{dm}}(a_{i})\text{\ensuremath{\Omega}}_{b0}\left(3c_{s,\mathrm{dm}}^{2}(a_{i})+1\right)} \, .
\end{equation}
In the next section, we solve Eqs. \eqref{delta_b_prime}-\eqref{theta_dm_prime} numerically and use the results to calculate the growth factor in our scenario.

\section{Growth of Structures}
\label{sec:growth_factor}

One of the most significant tests for our model lies in its predictions for the growth of cosmic structures. The quadratic EoS not only alters the background expansion but also introduces a non-zero, scale-dependent sound speed which actively suppresses the growth of perturbations on small scales. To assess the validity of our cosmological model in the linear regime of perturbations, we aim to study the growth of structures in this regime. To do so, we numerically integrate the full coupled system of perturbation equations \eqref{delta_b_prime}-\eqref{theta_dm_prime}, by implementing the initial conditions \eqref{delta_b_i}-\eqref{theta_dm_i}.

Here, we introduce the growth rate, which is related to the matter density contrast as
\begin{equation}
\label{eq:growth_rate}
f(a)=\frac{d\ln\delta_{m}}{d\ln a} \, .
\end{equation}
This quantity is related to the growth factor, which is an observable quantity and is defined by
\begin{equation}
\label{fsigma8_z}
f\sigma_{8}(z)=\frac{\sigma_{8}(z=0)}{\delta_{m}(z=0)}\:\left(a\:\frac{d\delta_{m}}{da}\right) \, .
\end{equation}
Here, $\sigma_8(z)$ is the root mean square matter fluctuation in spheres of radius $8h^{-1}\text{Mpc}$, and it is computed from the linear power spectrum $P(k,z)$ by using
\begin{equation}
\sigma_8^2(z) = \int_0^\infty \frac{k^2 dk}{2\pi^2} P(k,z) W^2(kR), \quad R = 8h^{-1}\text{Mpc} \, ,
\label{eq:sigma8_def}
\end{equation}
where $W(kR)$ is the Fourier transform of a spherical top-hat filter. The power spectrum is normalized to its CMB-inferred value at $z = 0$.

\begin{figure}[ht]
\centering
\includegraphics[width=12 cm]{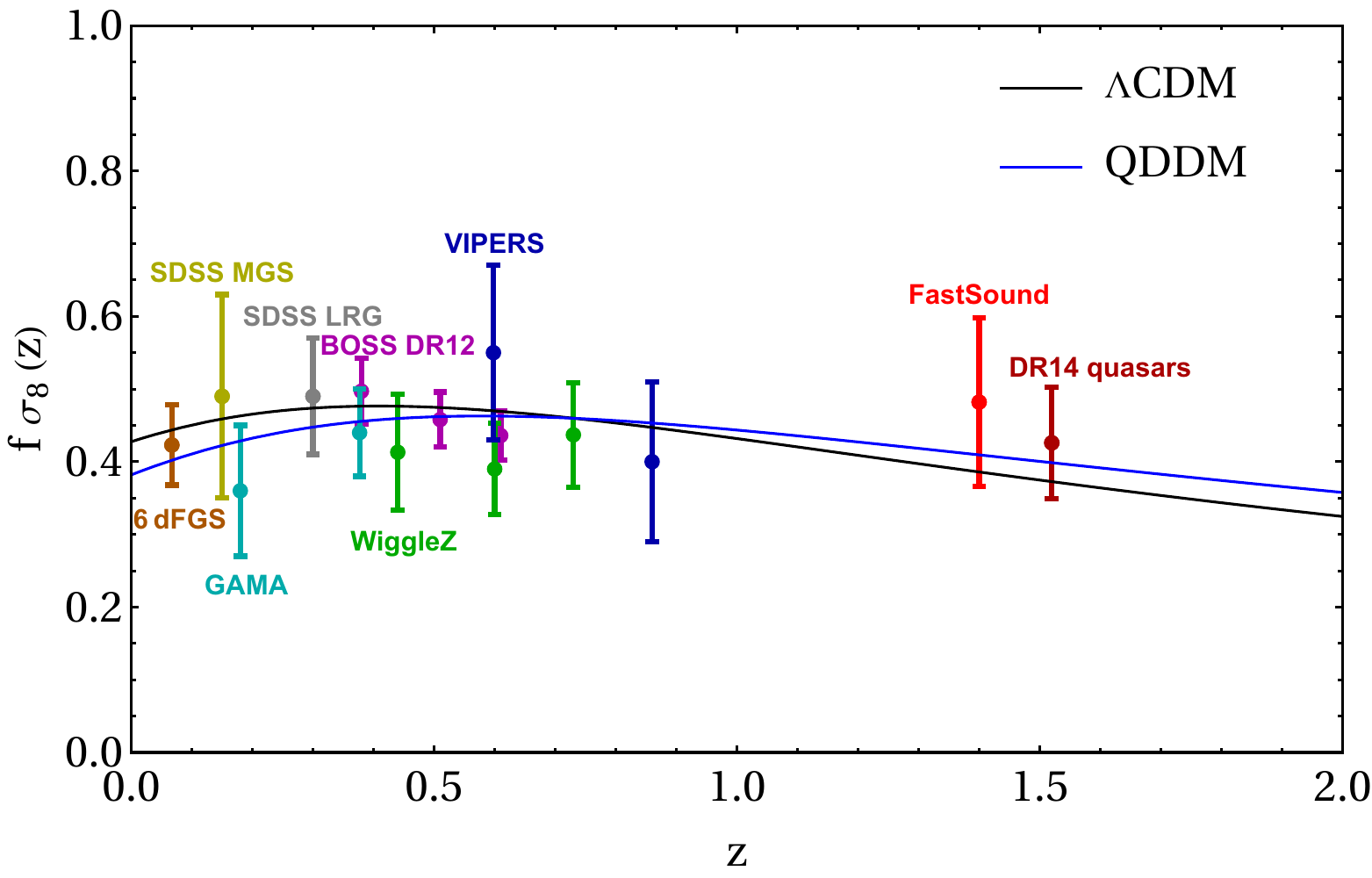}
\caption{\footnotesize Diagram of the growth factor for QDDM with $\{\alpha,\beta\}=\{2\times10^{-4},-10^{-5}\}$ (blue curve) compared to the $\Lambda$CDM result (black curve). To this plot we set $\sigma_{8}(z=0)=0.812$. The values of the other cosmological parameters are the same as those in Fig. \ref{fig:H}.}
\label{fig:fsigma8}
\end{figure}

Figure \ref{fig:fsigma8} shows the evolution of the observable growth factor $f\sigma_8(z)$ for our model with $\beta > 0$ compared to the typical $\Lambda$CDM expectation. The model predicts a lower amplitude of structure growth ($f\sigma_8$) at low redshifts ($z < 1-2$). This suppression arises from two interconnected effects:
\begin{enumerate}
\item \textbf{Modified Background:} The altered expansion history $H(z)$ changes the rate at which perturbations can grow. See Fig. \ref{fig:H} for more details.
\item \textbf{Pressure Support:} The effective sound speed $c_{s\,,\rm dm}^2$ \eqref{eq:cs2_pert} introduces pressure waves that counteract gravitational collapse, especially within the horizon. The term $\frac{k^{2}c_{s,\mathrm{dm}}^{2}}{a^{3}H^{2}\left(w_{\mathrm{dm}}+1\right)}\text{\ensuremath{\delta}}_{\mathrm{dm}}$ in the perturbation equations leads to a scale-dependent suppression, damping small-scale (large-$k$) modes more efficiently.
\end{enumerate}
This suppression of $f\sigma_8(z)$ is a hallmark of models with non-vanishing dark matter pressure. A value of $\beta < 0$ would have the opposite effect, potentially enhancing growth, though such models may face constraints from the observed matter power spectrum and CMB anisotropies.

\section{Conclusions}
\label{sec:conclusions}

In this work, we have introduced a phenomenological model for the DM component characterized by a quadratic equation of state \eqref{eq:main_eos}. This model represents a minimal yet powerful extension to the standard cosmological paradigm, moving beyond the simplistic assumption of perfectly cold, pressureless DM.

We compared the predictions of our model against the standard $\Lambda$CDM cosmology. We focus on key observational quantities: the background expansion history, the evolution of the dark matter density, and the growth rate of large-scale structure. This comparison highlights the unique phenomenological signatures imprinted by the quadratic term in the dark matter EoS.

Our analysis demonstrates that this model has profound and testable consequences for both the global expansion history of the Universe and the growth of cosmic structure. We derived the key equations governing the background evolution and the linear perturbations, revealing a rich phenomenology controlled primarily by the parameter $\beta$, which encapsulates the strength of the quadratic term.

Our principal conclusions are as follows:

\begin{itemize}

\item \textbf{Amplified Background Effects:} The proposed model introduces a new redshift-dependent deviation in the background expansion history. The quadratic term, significant only at high densities, introduces a redshift-dependent deviation in the expansion rate $H(z)$ (Fig. \ref{fig:H}). A tiny positive $\beta$ drives a faster expansion at high-$z$, while a tiny negative value slows a slower Hubble rate. This offers a potential pathway to address early universe-related tensions.

\item \textbf{Non-Standard Clustering:} It predicts a non-standard evolution of the DM density parameter. The modified expansion directly alters the evolution of the DM density parameter $\Omega_{\rm dm}(z)$ (Fig. \ref{fig:Omega}), changing the matter budget during the epoch of structure formation in a way that could be probed by combined CMB and LSS data.

\item \textbf{Suppressed Structure Growth:} The most significant signature is a suppression of the growth of structure, quantified by $f\sigma_8(z)$ at low redshifts (Fig. \ref{fig:fsigma8}). This arises from a dual mechanism: a modified background expansion and a scale-dependent pressure support ($c_{s,\mathrm{dm}}^2$) from the quadratic EoS. This provides a natural, physics-based mechanism to reduce the amplitude of late-time matter fluctuations, potentially resolving the $S_8$ tension.

\end{itemize}

The extent of these deviations is controlled by the parameter $\beta$. In the limit $\beta \rightarrow 0$, all our expressions reduce to the standard $\Lambda$CDM case, as required. In the upcoming work, we will confront these predictions with a suite of cosmological observations to place quantitative constraints on the allowed values of $\alpha$ and $\beta$.

The proposed model thus provides a unified framework that is consistent with standard late-time cosmology while offering new mechanisms to alter the Universe's early evolution and its structural growth history. The fact that such dramatic effects can stem from such a tiny fundamental parameter is a remarkable property of this model and points to a universe whose large-scale state is exquisitely sensitive to the precise microphysical nature of its dark components.

This work establishes the foundational theory and highlights the promising phenomenology of the QDDM model. The logical next step is to confront these predictions with a robust statistical analysis using a full suite of cosmological data. This includes fitting the model against the \textit{Planck} CMB power spectra, BAO measurements, SNIa distances, direct $H(z)$ constraints, and growth factor ($f\sigma_8$) data from redshift-space distortions. Such a Markov Chain Monte Carlo (MCMC) analysis will place precise quantitative constraints on the parameters $\alpha$ and $\beta$, determining if a non-zero quadratic EoS for dark matter is not just possible, but preferred by the data. Ultimately, this model opens a new avenue in the quest to understand the nature of dark matter, suggesting that its properties may be more complex than traditionally assumed, and that these complexities may hold the key to resolving the outstanding tensions in modern cosmology.


\bibliography{QDDM_JOFPA}

\end{document}